\documentclass[pre,showpacs,twocolumn]{revtex4}

\usepackage{graphicx}  
\usepackage{dcolumn}   
\usepackage{bm}        
\usepackage{subfigure}

\newcommand{\beq}{\begin{equation}}
\newcommand{\eeq}{\end{equation}}

\renewcommand{\phi}{\varphi}

\begin{document}

\title{Fingering instabilities of a reactive micellar interface
}


\author{Thomas Podgorski$^{1,2}$} 

\author{Michael C. Sostarecz$^1$} 
\thanks{Present address: Dept. of Physics, Monmouth College, Monmouth, IL.}

\author{Sylvain Zorman$^2$}

\author{Andrew Belmonte$^1$}
\thanks{Also at Harvard School of Engineering \& Applied Sciences, Harvard University, Cambridge, MA.}

\affiliation {$^1$W.~G.~Pritchard Laboratories, Department of Mathematics,\\
Penn State University, University Park, PA 16802, USA\\
$^2$Laboratoire de Spectrom\'etrie Physique,~CNRS--Universit\'e Joseph Fourier, Grenoble,\\ 38402 Saint Martin d'H\`eres, FRANCE}


\date{May 31, 2007}

\begin{abstract}
We present an experimental study of the fingering patterns in a Hele-Shaw cell, occurring when a gel-like material forms at
the interface between aqueous solutions of a cationic surfactant (cetyltrimethylammonium bromide) and an organic salt (salicylic acid), two solutions known to form a highly elastic wormlike micellar fluid when mixed homogeneously. A variety of fingering instabilities are observed, depending on the velocity of the front (the injection rate), and on which fluid is injected into which. We have found a regime of non-confined stationary or wavy fingers for which width selection seems to occur without the presence of bounding walls, unlike the Saffman-Taylor experiment. Qualitatively, some of our observations share common mechanisms with instabilities of cooling lava flows or growing biofilms. 
\end{abstract}

\pacs{83.60.Wc, 47.20.Gv, 83.80.Jx, 47.20.Ma}

\maketitle

\section{Introduction}

The classic instability of hydrodynamic fingering occurs when a fluid of a certain viscosity is injected into a more viscous fluid between closely-spaced parallel plates \cite{pelce}; this Saffman-Taylor instability \cite{saffman58} has been widely studied, and has had many variants, including fingering in polymer fluids, foams and gels \cite{nitt85,park94,lindner00,fast01,puff02}. 
 The morphology of the instability can also be influenced by anisotropy \cite{horvath87}, wetting \cite{alvarez} or modifications of the surface tension or density contrast between fluids due to a chemical reaction \cite{dewit99,dewit01}.  Hydrodynamic fingering is one of a wider class of instabilities occurring when one material is injected into another, or grows from a chemical or biological process; other such systems include smoldering flame fronts \cite{zik98}, filamentary microorganisms \cite{alain}, silica or iron tubes forming around metal salt solutions \cite{steinbock03,stone04}, electrochemical deposition of metals \cite{schroeter}, growing biofilms \cite{dockery01,rey01}, 
or lava flows \cite{
griffiths00}. In many of these cases, the interface itself is defined by a reaction or a solidification which changes the material properties, and is driven by the expanding growth of the interior.  The patterns in these systems result from a complex interaction of reaction, diffusion and mechanical effects in association with a particular rheology or elasticity of the interfacial zone.

Here we present a new type of pattern-forming instability, where fingering is linked to a rheological change - a gelling - due to a reaction which produces a viscoelastic micellar medium at the interface between two 
aqueous solutions
of identical viscosities in a Hele-Shaw cell \cite{BAPS05}. The micellar gel-like medium is formed by reaction-diffusion at the interface between 
two otherwise ordinary, 
miscible
water-like Newtonian fluids of identical viscosities. 
{It is worth mentioning that we use the word {\it interface} in a more general sense than as immiscible, sharp boundary. Here, the interface is a region which thickens in time and shows no discontinuity on a macroscopic scale.}
One distinguishing aspect of our system is that the fingering occurs independent of which fluid is injected into which, making this a true interfacial instability. In this paper we present our first exploration of this instability by varying fluid characteristics (concentrations) as well as the flow rate.

The fluids we study are aqueous solutions of the cationic surfactant cetyltrimethylammonium bromide (CTAB), and the organic salt sodium salycilate (NaSal), which form a strongly viscoelastic micellar material when brought into contact.  It is well known that a viscoelastic fluid is produced by the assembly of surfactants into long wormlike micelles, driven at low volume fractions by the mediating presence of certain slightly hydrophobic organic counterions \cite{larson,rehage91}.  The essential difference between these {\it wormlike micellar fluids} and more standard polymer fluids is that these aggregates are in a dynamic equilibrium with free surfactants in solution due to breaking and reforming processes \cite{israel}.

Much of the work on wormlike micellar fluids has focused on their unusual properties, either rheologically \cite{shikata88,grand,lerouge00, liuandpine} or hydrodynamically \cite{smolka03,handzy04,chen04,ballesta05}; there 
has been less study of the reaction process between the surfactant and 
the salt which gives rise to the wormlike micelles \cite{lin94}.  In 
fact, the simple act of preparing these fluids presents a startling 
phenomenon: two dilute solutions with essentially the physical 
properties of water become, upon combination, a strongly elastic 
fluid.

\section{Experimental Setup}

The experiment consists of injecting one of the solutions into a Hele-Shaw cell previously filled with the other solution.  
Two cells where used in two different laboratories.
Cell 1 (PSU) is made of two 
$9.5$~mm-thick square glass plates, 15 cm $\times$ 15 cm.  
Cell 2 (UJF) is made of two 10~mm thick glass disks, 20~cm in diameter.
A gap $b = 0.6$ mm between the two plates is fixed by brass shims. A light box below the cell provided a nearly uniform illumination (light box from Schott-Fostec for cell 1, a4 electro-luminescent panel from Selectronic for cell 2), and the pattern growth is observed from above by a COHU 4912 video camera connected to a Macintosh computer.

In most experiments, the cell is initially filled with a solution of NaSal (Sigma-Aldrich) of a given concentration, {after thorough cleaning of the glass plates so that the system is completely prewetted by the solution with no bubbles or wetting defects}.  A small amount of blue (McCormick) or green (Vahin\'e) food coloring is added to this solution for purposes of visualization.  The injected surfactant solution (CTAB, Sigma-Aldrich) is uncolored. In all experiments, both solutions have the same concentration, ranging from 25 to 50 mM. Injection is made through plastic tubing at the center of the top plate of the Hele-Shaw cell by a KDS100 syringe pump 
at a constant flow rate $Q$, which ranges from 0 to 1000~ml/h 
($0.278$~ml/s). In some experiments, the opposite arrangement of fluids was tested: NaSal solution injected into CTAB.

Two kinds of experiments were run: i) radial injection experiments, where the fluid is injected in an axisymmetric fashion in the Hele-Shaw cell, and ii) linear experiments where the initial direction of flow was imposed thanks to a U-shaped shim placed around the injection hole, aimed at studying isolated fingers.

\section{Experimental Results}

\subsection{Radial geometry}

\begin{figure}
\centering
{
\includegraphics[width=8cm]{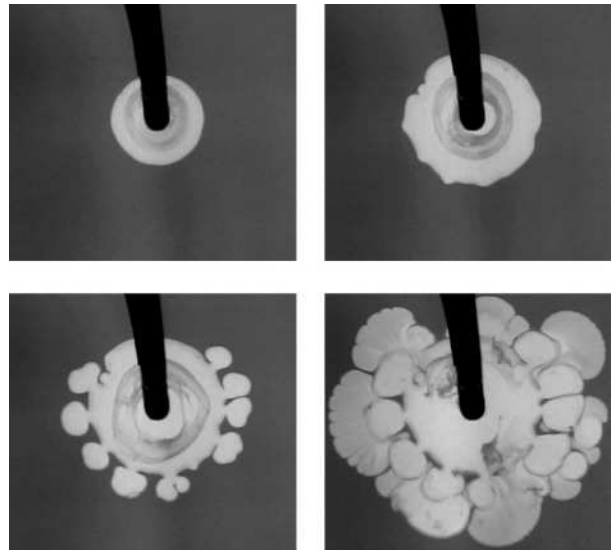} 
}

\caption{The onset of the fingering instability for an 
expanding circular front driven by an injection rate of  
150 ml/hr, 30 mM CTAB solution injected into 30 mM NaSal solution:
$t= 10.0$ s,
$t= 20.0$ s,
$t=28.8$ s,
$t=64.7$ s. Scale: picture width is 8~cm.
}

\label{f-instab} 
\end{figure}

\begin{figure}
\centering
{
\includegraphics[width=8cm]{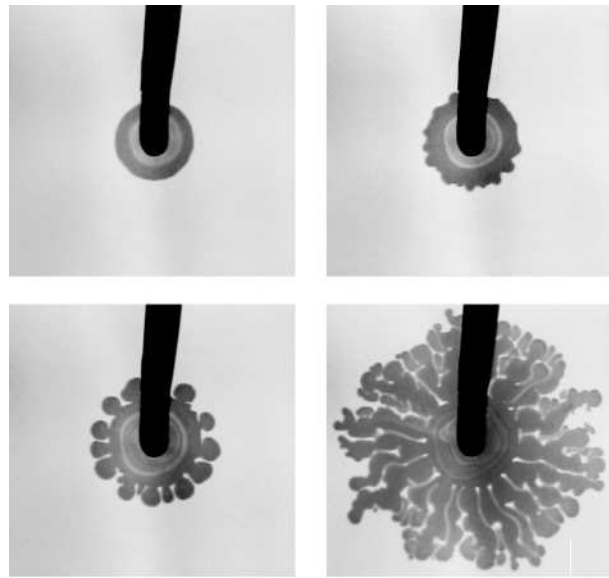}
} 


\caption{Fingering instability under the same conditions as Fig.~\ref{f-instab}, but with the opposite arrangement of fluids - 30 mM NaSal solution injected into 30 mM CTAB solution:
$t= 8.2$ s,
$t= 11.0$ s,
$t=16.3$ s,
$t=54.8$ s. Scale: picture width is 8~cm
}

\label{f-instab2} 
\end{figure}

The two solutions of CTAB and NaSal, at equimolar concentrations 
ranging from 25 to 50 mM are completely Newtonian before they are brought into contact with each other, each with a viscosity indistinguishable from that of pure water.  Consequently, in the absence of the ``wormlike 
micellar reaction'' at the interface, a Saffman-Taylor experiment 
would exhibit a completely stable radially expanding circle.  
Instead, as the front moves away from the injection point and the injected volume increases, a sequence of growth 
regimes takes place. {In addition, both fluids are miscible with no surface tension, and no wetting force exists at the moving front. One can therefore neglect wettability effects that play an important role in the Saffman-Taylor experiment with immiscible fluids \cite{alvarez}.}

We begin by discussing the injection of the surfactant solution (CTAB) into the solution with the organic counter-ion (NaSal).
A stable circular expansion is observed in our experiment at high injection speeds, which occur close to the injection point due to the constant flow-rate condition. In that case, when the radius $R$ of the expanding CTAB solution is small, or equivalently when the speed $\mathrm{d}R/\mathrm{d}t$ is large, the gelling reaction occurs at a slower timescale than the stretch rate of the gelled membrane which remains thin and offers no resistance to flow: the boundary between fluids is stable and isotropic.  The gel at the interface is probably very thin and dilute, and does not break or split (Fig.~\ref{f-instab}a).

When $R$ reaches some critical value which depends on flow-rate and concentrations, perturbations are observed 
on the smooth front (Fig.~\ref{f-instab}b).
The result of this instability is that the 
constant flow rate $Q$ focuses into the perturbations, which become 
more ``active'' and bulge out into mushroom shapes. Meanwhile, the regions between the bulges slow down, and appear to harden.  Subsequently, each individual mushroom spreads almost isotropically, albeit at a slower rate than the initial front, as each is effectively fed with a reduced flow rate (e.g.~about $Q/10$ in Fig.~\ref{f-instab}c). The appearance of these mushroom shapes around the perimeter bear a striking resemblance to the ``breakout'' of a lava flow \cite{griffiths00}.

\begin{figure}
\centerline{
\includegraphics[width=6.5cm]{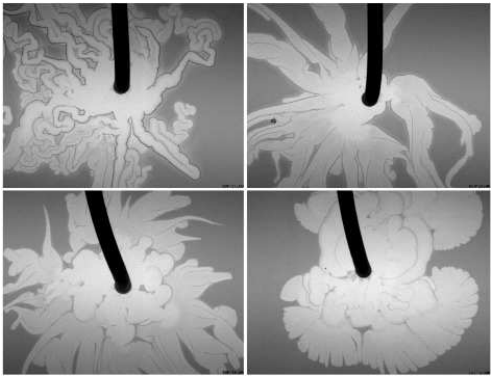} 
}

\caption{
Fingering patterns for different flow rates after the same total volume (4~ml) of 50mM CTAB has been injected into 50 mM NaSal:
a) 20 ml/h; 
b) 50 ml/h; c) 100 ml/h; d) 200 ml/h.
The dark borders are the gelled interfaces, which bound the interior, flowing conduit of fresh fluid which drives the further growth of the pattern. Scale: picture width is 11~cm.}
\label{f-morph}
\end{figure}

A precise determination of the onset of the first instability leading to the mushroom pattern can be made by measuring the radius of the circle circumscribing the growing front as a function of time; when the circular front becomes unstable to fingerlike protrusions, the growth rate of the gel front (and the entire pattern) goes through a rapid increase.  
We find an initial 
quantitative agreement with the constant flow rate law $R(t) = \sqrt{Qt 
/\pi b}$, where $Q$ is the imposed flow rate and $b$ is the 
gap width.  As the gel front slows, an instability appears nearly 
simultaneously around the circular front.  These perturbations grow, and take on the appearance of mushrooms, as shown in 
Fig.~\ref{f-instab}.  These new fronts, which have focused the flow 
from the center, move at a faster rate than the front just before the 
instability.  Eventually these fronts slow down as the pattern expands 
in size, and a second generation of fingers breaks from the pattern. 


Surprisingly, we observe the same instability to the mushroom pattern by performing the inverse experiment: injecting a 30 mM NaSal solution at the same rate into a 30 mM CTAB solution (Figure \ref{f-instab2}). The fact that the instability occurs independent of which fluid is injected into which distinguishes this from Saffman-Taylor and other related fingering instabilities, and shows that this fingering 
is truly driven by an interfacial instability.
Nevertheless, the critical radius at which the circular front becomes unstable is significantly smaller as can be seen when comparing Figures \ref{f-instab} and \ref{f-instab2}. This may be due to different diffusion coefficients of CTAB and Salicylate, or different compositions of the micellar media formed when injecting the surfactant into the salt or the opposite arrangement.

In the mushroom instability, the interfacial membrane between the two solutions appears to break simultaneously at a number of points along the perimeter. Bulges of fluid protrude, and the gel-like material forms at the new interfaces.  When this becomes more solid, it resists further smooth growth, but under the imposed conditions of constant injection rate, the pressure of the internal fluid will increase until 
a rupture of the gel membrane occurs again, and the process of 
fingering repeats.

\begin{figure}
\centerline{
\includegraphics[width=7.3cm]{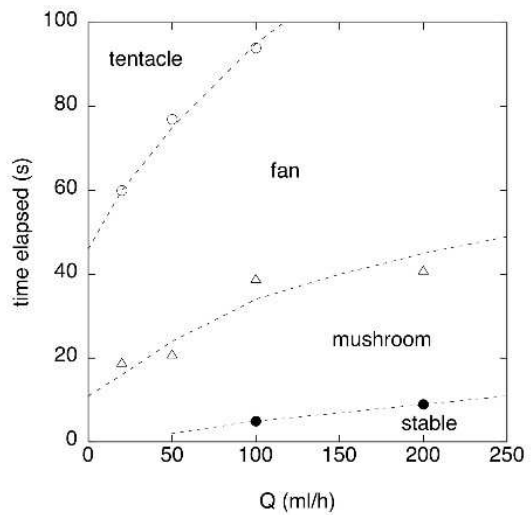}  
}

\caption{Phase diagram for the injection of 50 mM CTAB into 50 mM NaSal: Time at which successive regimes are triggered vs. flow rate. {$(\bullet)$: onset of mushroom growth, $(\triangle)$: onset of fans, $(\circ)$: onset of tentacle growth. Dashed lines are guidelines.}
}
\label{phasediag}
\end{figure}

Close observation of the interface between the two fluids reveals that the membrane is thickening with time, evidently due to the further reaction of surfactant and organic counterion diffusing through the material already formed. 
This thickening is made obvious by the slightly darker 
color of the material at the interface.  It has also been confirmed by a `post-mortem' test after the experiment: when the glass plates are separated, that the darker material is indeed gel-like, and clings to the glass while the two fluids rapidly flow away as the cell is emptied. This is most likely the reason that the membrane elasticity and resistance to flow increase with time. When some material threshold is reached \cite{gladdenPRL}, a fracture occurs in the membrane and fresh fluid leaks and forms the next generation of mushrooms.  However, the relatively well-defined wavelength of the mushroom pattern and the simultaneity of finger appearance suggests an underlying mechanism based on gel characteristics (thickness, concentration) prior to fracture.




Note that the subsequent evolution of the patterns in Figs.~1 and 2 differs for the two injection arrangements: while the experiment injecting CTAB solution into NaSal solution leads to other instabilities (described next), the other arrangement is simpler. The same process of expansion, hardening, fracture, and break-out seems to repeat as the pattern expands, without any other morphologies.

\begin{figure}
\centering
{\includegraphics[width=7.8cm]{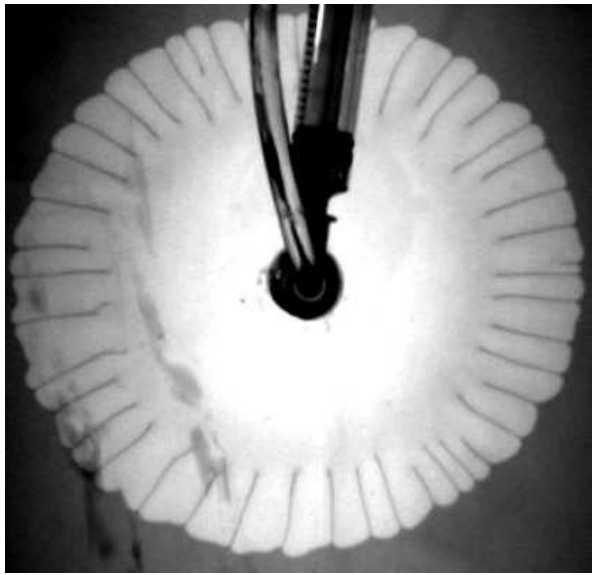}} 

\caption{\label{regfing} A regular pattern of fingers in the fan state, obtained when injecting 35 mM CTAB into 35 mM NaSal, $Q = 600$ ml/hr. Scale: picture width is 10 cm.}

\end{figure}  

We have identified two other instabilities which occur at later times as the radial pattern expands. After a few generations of mushroom patterns, we observe a transition to a fan-like pattern of contiguous wide fingers, giving the growing front a sort of flower-like appearance. 
The fingers emerging from these fans eventually become narrower ``tentacles'', which then undergo a curling instability and start to meander.  These tentacles seem to prefer growing along an existing structure rather than into fresh fluid.  
By changing the injection rate $Q$, the different growth regimes can be shifted both radially and in time: at higher injection rate, a given pattern 
instability is found to occur at larger distances from the center 
while at lower flow rates, the isotropic regime is almost invisible 
and the first observed pattern can be mushrooms or even multiple tentacles diverging from the injection point (see fig. \ref{f-morph}). The way this sequence of instabilities and patterns takes place as a function of flow rate is summarized in the schematic phase diagram of fig. \ref{phasediag} for 50mM concentrations.

While some regimes can be skipped by lowering the flow rate, which effectively  shrinks patterns radially down to the center, there are also ranges of concentrations and flow rates for which the fan instability is the first one to occur, directly after a stable circular front has grown to a finite distance to the center, as shown in Fig.~\ref{regfing} for $Q = 600$ ml/h and 35 mM concentration.

The second pattern observed in this experiment (``fans'') clearly involves an instability with a well defined wavelength (figs. \ref{f-morph}(d) and \ref{regfing}); while it appears to be similar to the first instability, the wavelength is much smaller.  This underlines a major influence of the radial expansion: a consequence of interface elongation is an azimuthal tension in the gel membrane which stabilizes the shortest wavelengths, in a similar fashion to surface tension in the classic Saffman-Taylor instability \cite{pelce}.  As the radius of the injected fluid increases, membrane stretch rate decreases and the interface destabilizes.

\begin{figure}
\centering
{\includegraphics[width=8cm]{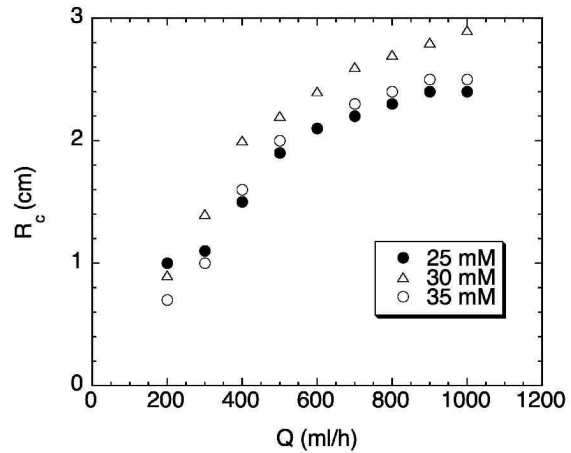}} 

\caption{\label{Rc} Critical radius of breakdown of axial symmetry vs. flow rate for three different CTAB / NaSal concentrations.}

\end{figure}  

More quantitative data on the onset of the first instability seen after the stable circular growth was obtained from the conditions when the stable circular growth stops and either a regular array of fingers characteristic of the fan state  (as in Fig.~\ref{regfing}), or a mushroom pattern (as in Fig.~\ref{f-instab}) appears. Figure \ref{Rc} shows the evolution of the critical radius at which the axial symmetry is lost vs. flow rate for three different concentrations. The critical radius increases with the flow rate: the gel membrane stretch rate being higher, it takes longer to become thick enough and trigger the instability. As far as concentration is concerned, one would intuitively expect that the stronger the concentration, the harder the gel. Thus the gel breaking should occur sooner for higher concentrations. However, an inversion is seen in  Fig.~\ref{Rc}: the curve for 25 mM is below others at high flow rate. This might be due to a different rheology at different concentrations (e.g. shear thinning occurring at different shear rates).

\begin{figure}
\centering
{
\includegraphics[width=8cm]{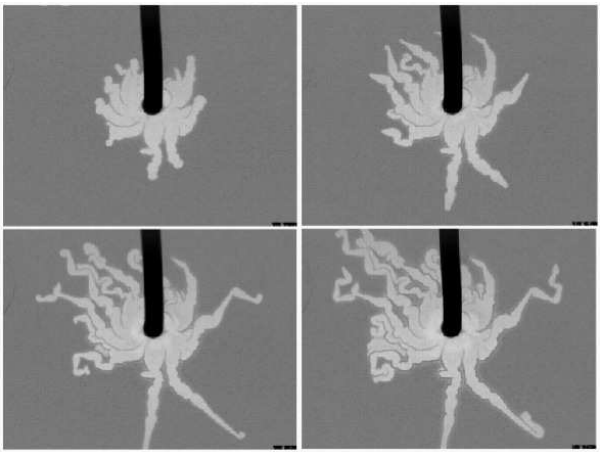}
}

\caption{\label{f-tentacle} Growth of the micellar gel pattern in the tentacle state, for the injection of a 50 mM CTAB solution into a 50 mM NaSal 
solution, at a flow rate $Q=20$~ml/h.  The width of each image is 11~cm.}

\end{figure}

The evolution of fingers in the tentacle state (Fig.~\ref{f-tentacle}) is significantly different than either the mushroom or fan state; the growth is very localized at the tip. The freshest fluid arrive at the tip from the interior of the finger,
 where it has not yet hardened to an immobile state, and is  advected to the sides by the feeding flow, where the interface solidifies.  We expect this instability to occur 
when the rate of creation of new interface exceeds the hardening time of the gel.



\begin{figure}
\centering
\includegraphics[width=8cm]{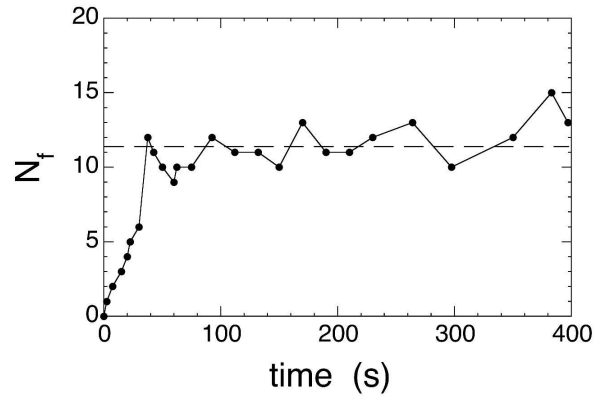}
\caption{\label{f-Nfingers} The total number of actively growing fingers $N_f$ in the tentacle growth state as a function of time (50 mM CTAB solution injected into 50 mM NaSal solution at 20 ml/hr).}

\end{figure}

A remarkable feature of these tentacles is their roughly constant width.  Since in this regime the average number of actively growing tentacles $N_f$ is typically constant (see Fig.~\ref{f-Nfingers}, which appears to have reached a steady state of $\sim$ 11.4 fingers), the velocity of the moving tip must on average also be constant.  Tentacles can therefore be seen as a state to which the system is attracted under certain conditions are reached: critical width (or wavelength of the ``fan'' instability) and critical velocity.

The stability of this regime can be qualitatively explained in the following way: a slight decrease of the finger width leads to a velocity increase because of the constant flow rate. The gel at the tip would then be weaker, which favors radial growth and tip widening. Conversely, a slight increase in finger width would decrease the finger velocity, which would give more time for the gel to harden. This would result in a greater resistance to finger widening.  Gel fracture would then occur and redirect the flow.  We focus on this finger regime in next section.

Curling structures appearing in the ultimate developments of tentacle growth, the tip favors motion along an existing membrane.  One mechanism for this could be that in these regions a depletion of the outer 
solution (NaSal) has occurred.  As a consequence, the gel forming there is weaker and poses less resistance to flow, resulting in an effective attraction between growing fingers.

\medskip

\subsection{Linear geometry - single finger growth}

While our initial observation of micellar interface fingering was made in a radial injection experiment, the geometry complicates the 
dynamics at a fixed flow rate, since the front speed decreases with radius as the pattern spreads out.  We therefore performed a second series of experiments in a linear geometry at LSP in Grenoble.

In order to investigate the remarkable finger growth regimes seen in the ultimate stages of radial growth, we made experiments focused on the study of this particular pattern.  A small device (U-shaped plate) was placed in the Hele-Shaw cell around the injection point to promote single finger growth in a certain direction, allowing precise control of the flow rate inside the finger.  The plate forms a small channel of width 3 mm and length about 2 cm.  Note that this channel serves {\it only} to set the initial condition: finger growth is observed once the fluid is coming out of the channel: the finger grows in an unconfined half-space.  When the flow rate $Q$ is in the correct range (approximately 0--3 ml/h), a single finger grows.  For $Q >$ 3 ml/hr, it becomes unstable and splits into two or more fingers.

\begin{figure}
\begin{center}
\includegraphics[width=8cm]{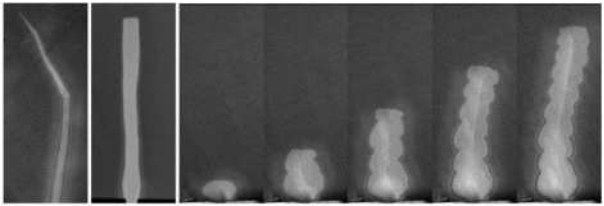}

\caption{\label{fingerpic} Linear finger growth regimes. Left to right: evanescent finger (concentration $c=30$ mM, flow rate $Q=1$ ml/h), steady finger ($c=40$ mM, $Q=2.5$ ml/h), wavy/oscillating finger growth ($c=30$ mM, $Q=1.5$ ml/h).  Each picture is 2 cm wide, and the time interval between wavy finger pictures is 150 s.}

\end{center}
\end{figure}

We observed different types of fingers, depending on flow rate and 
concentration see Fig.~\ref{fingerpic}.  For a given concentration, at 
low flow rates fingers are evanescent: their width decreases as the 
finger grows until the narrow tip opposes too much resistance to flow.  
Then the membrane breaks somewhere and a new evanescent finger grows. At intermediate flow rates (close to 2 ml/h), there exists a narrow range where steady growth occurs: fingers keep a constant width and the front moves at constant speed.  Above this range, one observes surprising oscillating fingers: the front periodically expands and 
contracts, modulating tip width.


Note that in both cases only the finger tip is active; the sides remain fixed, and only thicken slightly with time as discussed above.  Another remarkable feature is the tip shape: it is a straight front, perpendicular to the direction of motion, except for evanescent fingers at the end of their life.  This is completely unlike the rounded tips of most Saffman-Taylor fingers, and also the needle-like fingers seen in   fingering experiments on non-Newtonian associating polymer fluids \cite{ignes95} or colloidal pastes \cite{lemaire91}.


Moreover, the fact that the finger has a selected width while being far from any sidewalls is very different than the classic Saffman-Taylor case or the fan state described above, where the finger width is half of the channel width (without surface tension effects) \cite{pelce}; in a radial flow with far away boundaries the fingers expand and split \cite{homsy87,paterson81}.
This difference is a consequence of the rapid formation of a gel-like membrane that poses a strong resistance to flow, preventing finger widening. We suspect that in evanescent fingers the pressure and velocity are too low to prevent quick widening of the gel membrane toward the inside of the finger. The oscillation mechanism in wavy fingers is more mysterious. 
The oscillation mechanism
in wavy fingers is more mysterious. Given the fact that timescales involved in this process ($\sim 100$ s) could be of the same order of magnitude as rheological times of the micellar fluid, it may share a
common origin with the stick-slip instability suggested
by Puff  \emph{et al.} \cite{puff02}, however a more detailed characterization of the material will be needed to test the connection.

\section{Discussion and Conclusions}


We have presented here a new kind of fingering instability in a Hele-Shaw cell, defined by the interface between two solutions which together form wormlike micelles. The most striking aspect of these fingers is that the instability is completely determined by the properties of the interface. This is shown by the fact that essentially the same instability occurs independent of which solution is injected into which (in fact both solutions have the same viscosity). We have seen that there are some differences between these two patterns, which may arise from an asymmetry in diffusion coefficients of the two reacting species - it is very likely that the salicylate ion has a higher mobility in water than either a spherical micelle or a single surfactant molecule.

To our knowledge, the complex morphology of a growing, gelling 
interface has not previously been studied; however the early work by Hatschek on the gelling of a sinking drop of reactive fluid during vortex formation is similar in spirit \cite{hat19,thompson}. From our observations, many physical processes seem to influence pattern formation in this system: reaction-diffusion, flow and rheology of the viscoelastic micellar gel. The instabilities of such a system provide a new set of mechanisms which are of potential relevance to processes occurring in the growth of biological structures like bacterial biofilms.


Although our system is not explicitly biological, it includes many physical aspects of biofilms, such as the formation and growth of an elastic gel subjected to flow.  Our study sheds light on the purely physical instabilities to which such a system is susceptible.  The general problem of an expanding or growing gel finds an important application in certain biomedical infections, where the bacteria secrete a resistant ``biofilm'' matrix, which supports and protects them, often against disinfectants and antibiotics; an important example is \emph{Pseudmonas Aeruginosa}, responsible for many hospital infections \cite{costerton99}.  Our experimental results seem most closely related to other finger-like structures driven by dynamic growth or cooling processes, such as those produced by the rupture of the cooled crust of flowing lava  \cite{griffiths00}.  While much remains to be done to reach a quantitative understanding, the experiment we present here shares common qualitative features with many of these systems, and could be seen as a representative reaction-diffusion-advection system, displaying similar pattern formation but in a more convenient to controlled laboratory study than infectious biofilms or flowing lava.

\smallskip


AB acknowledges discussions with M. Clayton and J. P. Keener, and support from the Alfred P. Sloan Foundation, and the 
National Science Foundation (CAREER Award DMR-0094167).  TP and SZ wish to thank O. Pierre-Louis for helpful discussions and J. Sch\"{a}perklaus for experimental help.  TP also acknowledges financial support from DGA (French Ministry of Defense).

\end{document}